\documentclass[letterpaper]{article}

\usepackage{aaai2026}
\nocopyright

\usepackage{times}
\usepackage{helvet}
\usepackage{courier}
\usepackage[hyphens]{url}
\usepackage{graphicx}
\graphicspath{{figures/}{figure_drafts/}}
\def\UrlFont{\rm}
\usepackage{capt-of}
\usepackage{natbib}
\usepackage{booktabs}
\usepackage{multirow}
\usepackage{array}
\usepackage{algorithm}
\usepackage{algorithmic}
\usepackage{float}
\usepackage{dblfloatfix}
\usepackage{placeins}
\usepackage{amsmath}
\usepackage{amssymb}
\usepackage{xcolor}

\renewcommand{\topfraction}{0.98}
\renewcommand{\bottomfraction}{0.90}
\renewcommand{\dbltopfraction}{0.98}
\renewcommand{\textfraction}{0.02}
\renewcommand{\floatpagefraction}{0.80}
\renewcommand{\dblfloatpagefraction}{0.80}


\newcommand{\agentmeter}{\textsc{AgentMeter}}

\newcommand{\bodytablestyle}{%
  \footnotesize
  \setlength{\tabcolsep}{3.1pt}%
  \setlength{\aboverulesep}{0.30ex}%
  \setlength{\belowrulesep}{0.30ex}%
  \setlength{\belowcaptionskip}{2pt}%
  \renewcommand{\arraystretch}{1.03}%
}

\newcommand{\tablehead}[1]{\multicolumn{1}{c}{#1}}

\newcommand{\agentmeteropt}{\textsc{AgentMeter-Opt}}

\newcommand{\agentmeterteaserfigure}{%
\begin{figure}[!t]
  \centering
  \newcommand{\teaserpanelwidth}{0.97\linewidth}

  {\fontsize{8.5}{9.5}\selectfont\sffamily\bfseries
  \makebox[\linewidth][l]{(a) Motivation and components of AMS}}%
  \vspace{-1pt}

  \makebox[\linewidth][c]{%
    \includegraphics[
      width=\teaserpanelwidth,
      trim=22bp 28bp 20bp 24bp,
      clip
    ]{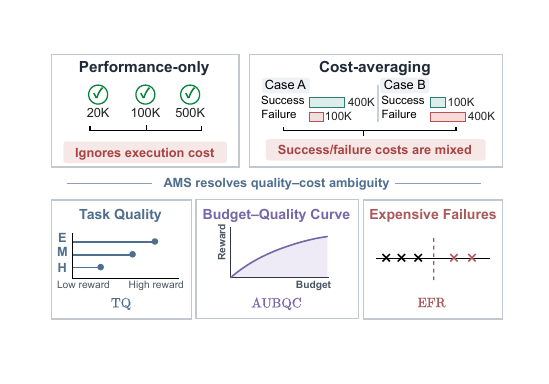}%
  }
  \vspace{-4pt}

  {\fontsize{8.5}{9.5}\selectfont\sffamily\bfseries
  \makebox[\linewidth][l]{%
    (b) AMS across LM--CLI pairs on the Core Subset}}%
  \vspace{-2pt}

  \makebox[\linewidth][c]{%
    \includegraphics[
      width=\teaserpanelwidth,
      trim=0bp 2bp 0bp 2bp,
      clip
    ]{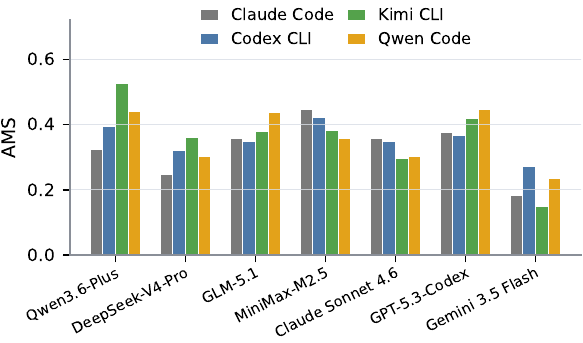}%
  }
  \vspace{-4pt}

  \caption{\textsc{AgentMeter} evaluates complete LM--CLI
configurations rather than models in isolation.
(a) AMS combines graded task quality, quality under calibrated cost
budgets, and costly zero-reward failures.
(b) AMS varies across CLIs for the same model; the best interface
therefore depends on the model.}
  \label{fig:agentmeter-overview}
\end{figure}
}

\title{
  Matching Matters: A Fair Quality--Efficiency Benchmark for\\
  Command-Line Agents
}

\author{
  Han Chi$^{1}$\equalcontrib,
  Jiaxin Qi$^{1}$\equalcontrib,
  Yan Cui$^{2}$,
  Baisheng Lai$^{1}$,
  Jianqiang Huang$^{1,2}$
}
\affiliations{
  $^{1}$Computer Network Information Center, Chinese Academy of Sciences\\
  $^{2}$Hangzhou Institute for Advanced Study, Chinese Academy of Sciences
}

\begin{document}

\maketitle

\begin{abstract}
Rapid advances in large language models have improved the task-solving
capabilities of command-line-interface (CLI)-based agents, whose CLIs
determine how models invoke tools, maintain interaction history, and
recover from failures. Consequently, effective matching between CLIs
and LLMs has become essential. However, existing agent benchmarks
largely emphasize success rate while overlooking practical objectives
such as cost and efficiency, as well as the selection of LM--CLI
combinations, all of which are critical in real-world deployment. We
therefore introduce \textsc{AgentMeter}, a quality--efficiency benchmark
with a new metric, the AgentMeter Score (AMS), that jointly
characterizes task quality, budget sensitivity, and resource-intensive
zero-reward execution, enabling a more complete assessment of deployed
LM--CLI pairs. Furthermore, collected task descriptions may
inadvertently favor LM--CLI pairs that are particularly compatible with
their wording and structure, causing evaluation results to reflect
description-specific advantages rather than general task-solving
capability. We therefore propose \textsc{AgentMeter-Opt}, a
trajectory-grounded optimization framework that constructs pair-adapted,
task-preserving description variants to build a fairer evaluation set
across LM--CLI pairs. Extensive experiments show that no CLI is
universally optimal across language models and that task success,
execution cost, and AMS identify different competitive configurations.
Results on \textsc{AgentMeter-Opt} further reveal that task-preserving
description changes affect LM--CLI pairs unevenly and can alter their
relative ordering across valid description conditions. Together,
\textsc{AgentMeter} and \textsc{AgentMeter-Opt} provide a practical
foundation for fair and deployment-relevant evaluation of command-line
agents.
\end{abstract}

\section{Introduction}
\agentmeterteaserfigure

Large language models (LMs) increasingly solve complex tasks through command-line interfaces (CLIs), which connect them to operating systems, files, executable programs, and development tools \citep{jimenez2024swebench,merrill2026terminalbench,huang2024dacode,liu2024agentbench}. A CLI forms part of the agent architecture: it determines which tools are exposed, how actions are encoded, what observations are returned, how history is retained, and how failures are handled. 
In practice, CLI implementations also incorporate model-specific choices in prompting and tool-use protocols. For example, Qwen Code and Kimi Code CLI provide integrations tailored to particular model families \citep{qwencode,kimicli}. Consequently, replacing the CLI while holding the model and task fixed can change the commands issued, the amount of backtracking, the execution cost, and the final outcome.

Most existing agent benchmarks evaluate each language model through a single fixed interface and summarize performance primarily in terms of task completion. 
Although this protocol enables controlled comparisons within a chosen
agent stack, it overlooks two deployment-relevant issues.
First, evaluation under a fixed interface obscures the effect of LM--CLI matching. Because CLIs differ in tool access, prompting, context management, and failure recovery, the same model may follow different execution trajectories and achieve different outcomes across interfaces. Conclusions drawn under one CLI may therefore not generalize to other LM--CLI configurations.
Second, task completion alone does not capture deployment efficiency. A desirable LM--CLI configuration should solve straightforward tasks with limited token use and terminate promptly when further progress is unlikely. Yet completion-only metrics assign the same credit to concise and costly successes, and the same penalty to early and prolonged failures.
Figure~\ref{fig:agentmeter-overview} illustrates these limitations. Panel~(a) shows that performance-only evaluation ignores execution cost, whereas average-cost evaluation mixes efficient successes with costly failures and can therefore obscure qualitatively different execution behavior. Panel~(b) further previews the substantial within-model variation across CLIs on our evaluation set.

To address these limitations, we introduce \textsc{AgentMeter}, an evaluation framework that treats the complete LM--CLI configuration as the evaluation unit while holding task specifications, execution environments, and evaluators fixed. At its core, the AgentMeter Score (AMS) captures three complementary aspects of deployment performance. First, it preserves graded task quality, distinguishing partial progress from executions that produce no useful result. Second, it measures how much quality is achieved under calibrated execution budgets, rewarding configurations that make useful progress available with less computation. Third, it penalizes costly zero-reward executions, distinguishing an early failure from a run that consumes substantial resources before returning no useful outcome. 
Because appropriate execution budgets depend on task demands, AMS calibrates these quantities within observed-effort strata rather than applying a single threshold to all tasks. 
We instantiate \textsc{AgentMeter} on 80 executable tasks from Terminal-Bench, SkillsBench, and DA-Code, and construct a 30-task Core Subset balanced across observed-effort strata for efficient evaluation of broad LM--CLI configuration grids \citep{merrill2026terminalbench,li2026skillsbench,huang2024dacode}.

However, a quality--efficiency metric alone does not ensure a fair comparison when each task is represented by only one collected description. The wording and structure of that description determine which requirements, artifact roles, and output constraints are made explicit, and may therefore favor LM--CLI pairs that are particularly compatible with its formulation. For example, when a required output format is left implicit, one pair may infer it immediately, whereas another may spend substantial resources inspecting artifacts. Measured performance can thus reflect compatibility with the collected description in addition to capability on the underlying task, consistent with prior findings that meaning-preserving changes in prompt wording and format can alter model behavior \citep{sclar2024promptformatting,zhu2024promptbench}.

To enable fair comparison, we introduce \textsc{AgentMeter-Opt}, a trajectory-grounded framework that deliberately optimizes a task-preserving description for every task--pair combination. 
Specifically, for each task and source LM--CLI pair, it uses the pair's original execution trajectory to identify description-related friction and generate targeted clarifications.
A task-preservation gate retains the original objective, artifact roles, required outputs, and success criterion while excluding hidden evaluation information and prescriptive solution steps. 
The retained description is selected to maximize performance on its source pair and is fixed before transfer to any other target. Crucially, this optimization is applied symmetrically: every source pair contributes one optimized description for every task. 
We then evaluate each target under Original, Self, Cross, and equal-weight Pooled conditions using the collected description and the resulting pair-optimized descriptions. 
In the Pooled condition, every target is evaluated on the same balanced set of descriptions with equal source weights. Because every source pair contributes equally, the comparison is less dependent on any single collected formulation and is therefore fairer across LM--CLI pairs.

Extensive experiments show that no CLI is uniformly best across language models, and that AMS identifies different leading configurations from completion- or cost-based metrics. Its components provide complementary ranking information, while the strong alignment between the Core and Full rankings (\(\rho=0.835\)) supports the Core Subset as a practical, efficient, and scalable setting for broad LM--CLI comparison. \textsc{AgentMeter-Opt} further shows that collected task descriptions are not neutral: pair-optimized descriptions improve Self performance, transfer unevenly across pairs, and alter rankings under Pooled evaluation, whereas public-only rewriting yields substantially smaller aggregate gains. Together, \textsc{AgentMeter} and \textsc{AgentMeter-Opt} provide a fair quality--efficiency benchmark for comparing LM--CLI configurations.

Our contributions are:

\begin{itemize}
\setlength{\itemsep}{0pt}
\setlength{\parsep}{0pt}
\setlength{\topsep}{2pt}




\item We identify two overlooked limitations in existing command-line agent evaluation: evaluating language models through a single fixed interface obscures LM--CLI matching effects, while completion-oriented metrics overlook deployment efficiency.

\item We introduce \textsc{AgentMeter}, an agent benchmark with a Core Subset for systematically and efficiently evaluating various LM--CLI configurations. We further propose the AgentMeter Score (AMS), which jointly captures graded task quality, quality under calibrated execution budgets, and costly zero-reward execution.

\item We introduce \textsc{AgentMeter-Opt}, a trajectory-grounded optimization method that constructs one pair-optimized and task-preserving description for every task--pair combination. By applying this optimization symmetrically and evaluating all targets on the same balanced description set, it reduces dependence on any single collected formulation and enables fairer comparison across LM--CLI configurations.

\end{itemize}







\begin{figure*}[!t]
  \centering
  \makebox[\textwidth][c]{%
    \includegraphics[
      width=\textwidth,
      trim=50bp 42bp 82bp 42bp,
      clip
    ]{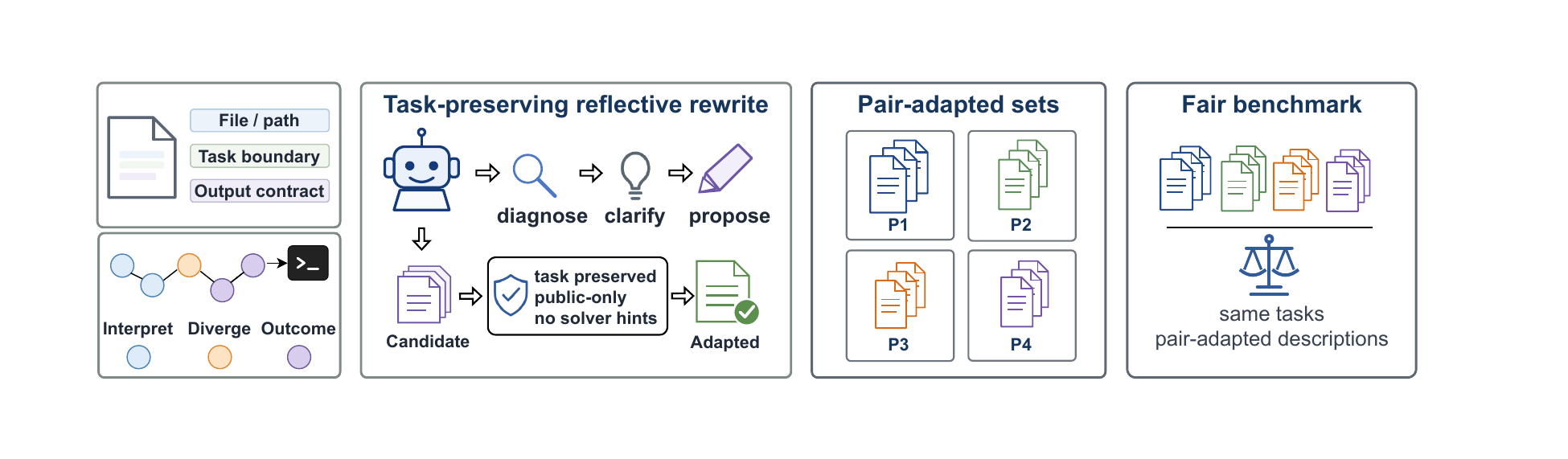}%
  }
  \caption{\textsc{AgentMeter-Opt} uses each source pair's Original
  trajectory to construct a task-preserving description. Target pairs
  are evaluated on the same tasks under Original, Self, Cross, and
  source-balanced Pooled conditions.}
  \label{fig:agentmeter-opt}
\end{figure*}

\section{Related Work}
\label{sec:related-work}

\noindent\textbf{Agent benchmarks.}
Agent benchmarks increasingly evaluate language models through
interaction with executable environments. SWE-bench and
Terminal-Bench study repository-level software engineering and
terminal tasks
\citep{jimenez2024swebench,merrill2026terminalbench}, while
SkillsBench and DA-Code cover structured skill use and multi-step data
analysis
\citep{li2026skillsbench,huang2024dacode}. AgentBench, GAIA,
WebArena, WorkArena, and OSWorld extend evaluation to general tool use,
web navigation, enterprise applications, and desktop interaction
\citep{liu2024agentbench,mialon2024gaia,zhou2024webarena,
drouin2024workarena,xie2024osworld}. These benchmarks provide
realistic tasks and execution-based evaluation, but model comparisons
are usually conducted through a fixed agent interface.
\textsc{AgentMeter} instead treats the complete LM--CLI pair as the
evaluation unit and compares multiple models and CLIs under the same
tasks, environments, and evaluators.

\noindent\textbf{Agent interfaces and evaluation metrics.}
Agent performance depends on more than final task completion.
AgentBoard measures intermediate trajectory progress, and
\textit{AI Agents That Matter} studies accuracy--cost trade-offs
\citep{ma2024agentboard,kapoor2025agentsmatter}. SWE-agent and
OpenHands further show that observations, action spaces, tool
protocols, and execution harnesses affect agent behavior
\citep{yang2024sweagent,wang2025openhands}. Practical systems such as
Qwen Code and Kimi Code CLI likewise differ in prompting, tool use,
context management, and terminal execution
\citep{qwencode,kimicli}. AMS complements these studies by
jointly measuring graded task quality, quality achieved within
calibrated budgets, and costly zero-reward execution when ranking
LM--CLI pairs.

\noindent\textbf{Prompt sensitivity and optimization.}
Meaning-preserving changes in wording, formatting, and prompt
templates can alter model outcomes
\citep{sclar2024promptformatting,zhu2024promptbench}. Automatic prompt
optimization methods such as APE, OPRO, ProTeGi, PromptBreeder, DSPy,
TextGrad, and GEPA use candidate search or execution feedback to
improve prompts and LM programs
\citep{zhou2023ape,yang2024opro,pryzant2023protegi,
fernando2024promptbreeder,khattab2024dspy,
yuksekgonul2025textgrad,agrawal2026gepa}; Reflexion similarly uses
prior experience to improve later executions
\citep{shinn2023reflexion}. These methods primarily optimize performance for a target system.
\textsc{AgentMeter-Opt} instead uses task-preserving optimization to
study evaluation sensitivity by constructing source-conditioned
descriptions and comparing LM--CLI pairs under Original, Self, Cross,
and equal-weight Pooled conditions. Each description is fixed before
transfer, and every target is evaluated over the same source-balanced
set. Thus, description variation becomes part of the evaluation
protocol rather than a target-side tuning advantage.

\section{Methodology}
\label{sec:method}

\subsection{Problem Formulation}
\label{sec:problem-formulation}

We treat the pairing of a language model and a command-line interface
as an LM--CLI configuration, denoted by $\pi$. Each task $i$ is
specified by its objective, input artifacts and their roles, required
outputs, execution environment, evaluator, and success criterion.
Together, these elements define the task semantics. Given a public task, executing configuration $\pi$ on task $i$ produces a trajectory $\tau_i(\pi)$, an evaluator reward
$R_i(\pi)\in[0,1]$, and a billable execution cost
$C_i(\pi)\geq 0$.




\subsection{AgentMeter Score}
\label{sec:ams}

We first collect the dataset as AgentMeter, which evaluates each LM--CLI pair under the original description of every task. Then, we propose AgentMeter Score
(AMS), which measures not only how much task quality a configuration achieves,
but also how efficiently it achieves that quality and how much it spends
on zero-reward executions. AMS combines three complementary
components: {{T}ask {Q}uality} (TQ), {cost-aware Area Under the
Budget--Quality Curve} (AUBQC), and {Expensive Failure Rate}
(EFR). 

Because execution costs are meaningful only relative to task demand,
AMS first estimates the observed effort of each task from a fixed
reference set of LM--CLI configurations, denoted by
$\Pi_{\mathrm{ref}}$. Throughout this subsection,
$\tau_i(\pi)$, $R_i(\pi)$, and $C_i(\pi)$ denote the trajectory,
reward, and cost obtained by configuration $\pi$ on task $i$ under its
collected public description. Let $\operatorname{Tok}(\tau)$ and
$\operatorname{Step}(\tau)$ denote the token usage and number of
execution steps recorded in trajectory $\tau$, respectively. We define
the observed-effort index of task $i$ as

\begin{equation}
\label{eq:effort-index}
{\small
\begin{aligned}
e_i
={}&
q\!\left(
\operatorname*{median}_{\pi\in\Pi_{\mathrm{ref}}}
\operatorname{Tok}\!\left(\tau_i(\pi)\right)
\right)
+
q\!\left(
\operatorname*{median}_{\pi\in\Pi_{\mathrm{ref}}}
\operatorname{Step}\!\left(\tau_i(\pi)\right)
\right)
\\[-1pt]
&+
q\!\left(
\mathbb{E}_{\pi}
\!\left[
\mathbf{1}\!\left(R_i(\pi)=0\right)
\right]
\right).
\end{aligned}
}
\end{equation}
where each median is taken across the reference configurations and
$q(\cdot)$ converts a task-level statistic into its percentile rank
among the reference tasks. The first term measures the typical token
demand, the second measures its typical interaction length,
and the third measures how frequently reference configurations obtain
zero reward. Aggregating their percentile ranks yields a common effort scale that reflects both resource demand and observed execution difficulty.

The effort index is used only to calibrate task-dependent cost scales
and does not directly enter AMS. 
We sort tasks by $e_i$ and partition
the ordered list into three nearly equal effort strata,
$\mathcal{S}\!=\!\{\text{Easy},\text{Medium},\text{Hard}\}$. 
For each
$s\in\mathcal{S}$, let $\mathcal{T}_s$ denote the corresponding task
set.

With these task groups fixed, AMS first measures the graded quality
achieved by a configuration. We define the task-quality score within
stratum $s$ as

\begin{equation}
\label{eq:task-quality}
S_{\mathrm{TQ},s}(\pi)
=
\frac{1}{|\mathcal{T}_s|}
\sum_{i\in\mathcal{T}_s}
R_i(\pi).
\end{equation}

This score averages the evaluator rewards across tasks in the stratum,
thereby preserving partial progress rather than reducing each execution
to a binary success or failure.

Task quality alone does not distinguish configurations that achieve the
same reward at different execution costs. To measure how much quality
is available under different cost constraints, let $\mathcal{B}_s$
denote a fixed grid of billable-cost budgets calibrated from
positive-reward reference executions in stratum $s$. We define the
cost-aware area under the budget--quality curve as

\begin{equation}
\label{eq:costaubqc}
A_{\mathrm{BQ},s}(\pi)
=
\frac{1}{|\mathcal{B}_s|\,|\mathcal{T}_s|}
\sum_{b\in\mathcal{B}_s}
\sum_{i\in\mathcal{T}_s}
R_i(\pi)\,
\mathbf{1}\!\left[C_i(\pi)\leq b\right].
\end{equation}

For each budget $b$, an execution contributes its evaluator reward only
when its cost does not exceed that budget. A configuration that achieves
the same reward at a lower cost therefore contributes under more budget
levels and attains a larger $A_{\mathrm{BQ},s}(\pi)$. This quantity is
the AUBQC component of AMS. Zero-reward executions, however,
contribute nothing at every budget level regardless of their cost,
which motivates the separate failure component introduced next.

Zero-reward executions require separate treatment because
$A_{\mathrm{BQ},s}(\pi)$ assigns them zero contribution regardless of
their cost. Let $\eta_s$ denote a fixed expensive-failure threshold
calibrated from zero-reward reference executions in stratum $s$. We
define the expensive failure rate as

\begin{equation}
\label{eq:expensive-failure}
R_{\mathrm{EFR},s}(\pi)
=
\frac{1}{|\mathcal{T}_s|}
\sum_{i\in\mathcal{T}_s}
\mathbf{1}\!\left[
R_i(\pi)=0
\;\land\;
C_i(\pi)>\eta_s
\right].
\end{equation}

This rate measures how often a configuration consumes unusually high
cost without obtaining any evaluator reward, thereby distinguishing
early failures from resource-intensive zero-reward executions.

Having defined the three components, we combine task quality and
budget-conditioned quality before accounting for expensive failures.
For a global mixing weight $\alpha$, the score within stratum
$s$ is

\begin{equation}
\label{eq:stratum-score}
S_s(\pi)
=
\left[
\alpha S_{\mathrm{TQ},s}(\pi)
+
(1\!-\!\alpha)A_{\mathrm{BQ},s}(\pi)
\right]
\left[
1-R_{\mathrm{EFR},s}(\pi)
\right].
\end{equation}

The weighted sum balances overall task quality against the quality
achieved within cost budgets, while the multiplicative term discounts
configurations that frequently incur expensive zero-reward executions.

Finally, AMS assigns equal importance to the three effort strata by
macro-averaging their scores:

\begin{equation}
\label{eq:ams}
\mathrm{AMS}(\pi)
=
\frac{1}{|\mathcal{S}|}
\sum_{s\in\mathcal{S}}
S_s(\pi).
\end{equation}

The calibration set, budget grids, failure thresholds, mixing weight,
and pricing rules are fixed before scoring and applied consistently
across all evaluated LM--CLI configurations.

\begin{table*}[!t]
\centering
\footnotesize

\setlength{\tabcolsep}{2.2pt}
\setlength{\aboverulesep}{0.28ex}
\setlength{\belowrulesep}{0.28ex}
\renewcommand{\arraystretch}{0.92}

\begin{tabular*}{0.93\textwidth}{
  @{\extracolsep{\fill}}
  l@{\hspace{3pt}}l
  *{9}{c}
  @{}
}
\toprule
\tablehead{Model}
& \tablehead{CLI}
& \tablehead{Pass$\uparrow$}
& \tablehead{Tier-P$\uparrow$}
& \tablehead{Tok/P$\downarrow$}
& \tablehead{USD/P$\downarrow$}
& \tablehead{TQ$\uparrow$}
& \tablehead{AUBQC$\uparrow$}
& \tablehead{EFR$\downarrow$}
& \tablehead{AMS$\uparrow$}
& \tablehead{Gap$\downarrow$} \\
\midrule

\multirow{4}{*}{Qwen3.6-Plus}
& Claude & 11 & 7/4/0 & 3.51M & 0.332
& 0.394 & 0.251 & 0.033 & 0.325 & 0.200 \\
& Codex & 12 & 8/3/1 & \textbf{0.48M} & \textbf{0.074}
& 0.417 & 0.357 & 0.000 & 0.393 & 0.132 \\
& Kimi & 16 & 8/6/2 & 1.71M & 0.198
& 0.573 & \textbf{0.453} & 0.000 & \textbf{0.525}
& \textbf{0.000} \\
& Qwen & 14 & 8/4/2 & 1.89M & 0.204
& 0.501 & 0.349 & 0.000 & 0.440 & 0.085 \\

\specialrule{0.3pt}{1.0pt}{1.0pt}

\multirow{4}{*}{GLM-5.1}
& Claude & 15 & 8/4/3 & 1.56M & 0.830
& 0.539 & 0.159 & 0.100 & 0.358 & 0.167 \\
& Codex & 14 & 8/5/1 & 1.56M & 0.667
& 0.487 & 0.196 & 0.067 & 0.348 & 0.177 \\
& Kimi & 16 & 9/5/2 & 1.30M & 0.608
& 0.559 & 0.214 & 0.133 & 0.380 & 0.145 \\
& Qwen & \textbf{18} & 8/6/4 & 1.56M & 0.633
& 0.621 & 0.239 & 0.100 & 0.437 & 0.088 \\

\specialrule{0.3pt}{1.0pt}{1.0pt}

\multirow{4}{*}{GPT-5.3-Codex}
& Claude & 16 & 7/5/4 & 0.62M & 0.335
& 0.557 & 0.221 & 0.100 & 0.375 & 0.150 \\
& Codex & 16 & 8/4/4 & 0.90M & 0.471
& 0.553 & 0.192 & 0.100 & 0.368 & 0.157 \\
& Kimi & 16 & 8/4/4 & 0.55M & 0.396
& 0.553 & 0.283 & 0.067 & 0.420 & 0.105 \\
& Qwen & 17 & 8/5/4 & 0.55M & 0.290
& 0.598 & 0.265 & 0.033 & 0.445 & 0.080 \\

\specialrule{0.3pt}{1.0pt}{1.0pt}

\multirow{4}{*}{DeepSeek-V4-Pro}
& Claude & 13 & 8/4/1 & 1.83M & 1.476
& 0.461 & 0.044 & 0.233 & 0.246 & 0.279 \\
& Codex & 14 & 8/5/1 & 1.84M & 1.298
& 0.502 & 0.103 & 0.133 & 0.320 & 0.205 \\
& Kimi & 16 & 8/5/3 & 1.01M & 0.860
& 0.561 & 0.189 & 0.133 & 0.362 & 0.163 \\
& Qwen & 14 & 8/5/1 & 1.39M & 0.954
& 0.514 & 0.123 & 0.167 & 0.304 & 0.222 \\

\specialrule{0.3pt}{1.0pt}{1.0pt}

\multirow{4}{*}{Claude Sonnet 4.6}
& Claude & 17 & 8/5/4 & 0.98M & 0.733
& 0.588 & 0.113 & 0.100 & 0.358 & 0.167 \\
& Codex & 17 & 7/6/4 & 1.04M & 0.808
& 0.607 & 0.134 & 0.167 & 0.348 & 0.177 \\
& Kimi & 15 & 7/5/3 & 0.67M & 1.074
& 0.511 & 0.053 & 0.100 & 0.295 & 0.230 \\
& Qwen & 16 & 8/5/3 & 1.67M & 2.477
& \textbf{0.627} & 0.061 & 0.267 & 0.303 & 0.222 \\

\specialrule{0.3pt}{1.0pt}{1.0pt}

\multirow{4}{*}{Gemini 3.5 Flash}
& Claude & 12 & 7/4/1 & 9.95M & 11.995
& 0.411 & 0.007 & 0.367 & 0.183 & 0.342 \\
& Codex & \textbf{18} & 8/5/5 & 3.15M & 2.013
& 0.620 & 0.027 & 0.300 & 0.273 & 0.252 \\
& Kimi & 9 & 7/1/1 & 3.45M & 1.978
& 0.311 & 0.007 & 0.233 & 0.150 & 0.375 \\
& Qwen & 15 & 7/4/4 & 3.45M & 1.963
& 0.521 & 0.029 & 0.267 & 0.236 & 0.289 \\

\specialrule{0.3pt}{1.0pt}{1.0pt}

\multirow{4}{*}{MiniMax-M2.5}
& Claude & 13 & 8/3/2 & 1.94M & 0.133
& 0.488 & 0.386 & 0.000 & 0.447 & 0.078 \\
& Codex & 12 & 5/5/2 & 1.19M & 0.103
& 0.452 & 0.378 & 0.000 & 0.422 & 0.103 \\
& Kimi & 11 & 8/3/0 & 1.99M & 0.145
& 0.398 & 0.362 & 0.000 & 0.383 & 0.142 \\
& Qwen & 11 & 7/3/1 & 2.76M & 0.179
& 0.391 & 0.304 & 0.000 & 0.356 & 0.169 \\

\bottomrule
\end{tabular*}

\caption{Core results for 28 LM--CLI configurations. Pass, cost per
success, and AMS identify different leaders. Tier-P reports
Easy/Medium/Hard passes; Gap is measured from the best AMS.}
\label{tab:core30-main-results}
\vspace{-3pt}
\end{table*}

\subsection{\textsc{AgentMeter-Opt}}
\label{sec:agentmeter-opt}

As we discussed, evaluating every LM--CLI
configuration under a single collected task description can make the
measured comparison depend on that description's wording and structure.
A description may expose requirements, artifact roles, or output
constraints in ways that favor configurations particularly compatible
with its formulation, thereby confounding task-solving capability with
description-specific compatibility. 

To reduce this dependence,
\textsc{AgentMeter-Opt} constructs task-preserving descriptions adapted
to different LM--CLI configurations and evaluates each configuration
under the resulting description set.
Let $\mathcal{P}\!=\!\{\pi_1,\ldots,\pi_K\}$ denote the fixed set of
LM--CLI configurations. For task $i$, let $d_i$ denote its collected description. For
each source configuration $\pi_k\in\mathcal{P}$,
\textsc{AgentMeter-Opt} constructs an optimized description
$d_{i,k}^{*}$ from $d_i$ and the corresponding source trajectory
$\tau_i(\pi_k;d_i)$.

\noindent\textbf{Task-preserving prompt optimization.}
Let $\mathcal{D}_{i,k}$
denote the space of reformulations conditioned on $d_i$ and
$\tau_i(\pi_k;d_i)$ that preserve the task semantics and task boundary, while introducing
neither hidden evaluator information, new task requirements, reference
answers, nor prescriptive solution steps. The optimized description is

\begin{equation}
\label{eq:description-optimization}
d_{i,k}^{*}
=
\operatorname*{arg\,max}_{\tilde d\in\mathcal{D}_{i,k}}
\left(
R_i(\pi_k;\tilde d),
-C_i(\pi_k;\tilde d)
\right),
\end{equation}
where the objective maximizes the evaluator reward achieved by
configuration $\pi_k$ and then favors lower billable cost among
descriptions with equal reward. The resulting description adapts the
task formulation to the source configuration without changing the
underlying task.

\noindent\textbf{A fair source-balanced evaluation set.}
The collected evaluation set associates each task with a single
description,
$\mathcal{E}=\{(i,d_i)\mid i\in\mathcal{T}\}$. Then,
\textsc{AgentMeter-Opt} expands it into

\begin{equation}
\label{eq:source-balanced-set}
\mathcal{E}^{*}
=
\left\{
(i,d_{i,k}^{*})
\mid
i\in\mathcal{T},\;
\pi_k\in\mathcal{P}
\right\}.
\end{equation}

Thus, each task is represented by one task-preserving description from
every source configuration. Every LM--CLI pair contributes equally, and every
configuration is evaluated on the same expanded set. This symmetric
construction preserves the underlying tasks while reducing the
influence of any particular description, enabling a fairer comparison
across LM--CLI configurations.

\section{Experiments}
\label{sec:experiments}

\subsection{Datasets}
\label{sec:datasets}

The Full Benchmark contains 80 executable command-line tasks from
SkillsBench, Terminal-Bench, and DA-Code
\citep{li2026skillsbench,merrill2026terminalbench,huang2024dacode}.
It provides the reference task distribution used for effort and cost
calibration. The 30-task Core Subset preserves coverage of the source
benchmarks and observed-effort strata while supporting the complete
\(7\times4\) LM--CLI grid, yielding 840 pair--task runs. We assess its fidelity using the 16 configurations with
complete Full-Benchmark coverage, comprising 1,280
Full-Benchmark pair--task runs in total.

For \textsc{AgentMeter-Opt}, we use a fixed 16-task subset spanning
5 Easy, 7 Medium, and 4 Hard tasks to study how task-preserving
description variants affect LM--CLI performance and ranking.

\subsection{Implementation Details}
\label{sec:implementation-details}

\noindent\textbf{Models and interfaces.}
We evaluate seven language models: Qwen3.6-Plus,
DeepSeek-V4-Pro, GLM-5.1, MiniMax-M2.5,
Claude Sonnet 4.6, GPT-5.3-Codex, and Gemini 3.5 Flash
\citep{qwen36plus,deepseekv4pro,glm51,minimaxm25,
claudesonnet46,gpt53codex,gemini35flash}.
Each model is paired with Claude Code, Codex CLI, Kimi Code CLI,
and Qwen Code
\citep{claudecode,codexcli,kimicli,qwencode}.
Exact provider model IDs, snapshots, inference settings, CLI releases
or commits, and evaluation dates are reported in the supplementary
material.

\noindent\textbf{Evaluation protocol.}
AMS is the primary metric. We additionally report pass count, tokens
per successful task, and USD per successful task. We set
\(\alpha=0.6\). Within each Full-Benchmark effort stratum, the
billable-cost budget grid is calibrated at the 20th, 40th, 60th,
80th, and 90th percentiles of positive-reward reference executions.
The expensive-failure threshold is the 90th percentile of zero-reward
execution cost in the same stratum. All calibration choices, including
the reference configurations, effort strata, budget grids, failure
thresholds, and pricing rules, are fixed in advance and shared across
all evaluated configurations. Unless otherwise stated, 95\% confidence
intervals use 10,000 task-level paired bootstrap resamples.

\noindent\textbf{Description construction.}
We use four LM--CLI pairs: Qwen3.6-Plus/Kimi Code CLI,
Qwen3.6-Plus/Codex CLI, DeepSeek-V4-Pro/Kimi Code CLI, and
GLM-5.1/Qwen Code. This selection includes both
same-model/different-CLI and different-model/same-CLI comparisons,
allowing us to examine description transfer across models and
interfaces. For each task--source pair, Qwen3.7-Max
\citep{qwen37max} diagnoses description-related friction from the
Original trajectory and generates candidate clarifications under a
shared prompt and decoding configuration.

Candidates are filtered and selected using the task-preservation and
source-side selection procedure described in the Methodology section.
Only task-preserving candidates are eligible; selection first considers
source-side reward and then billable cost, retaining Original when no
candidate improves reward or preserves reward at lower cost. The
selected description is frozen before target evaluation, so no
target-side trajectory, reward, or cost can affect selection or
fallback.

\noindent\textbf{Evaluation conditions and controls.}
Each target pair is evaluated under Original, Self, and Cross
conditions. The resulting matrix contains 64 Original, 64 Self, and
192 Cross pair--task executions, for 320 executed conditions in total.
Pooled assigns equal weight to the four source-conditioned outcomes
for each target, whereas Cross-only averages the three outcomes whose
descriptions were not derived from that target pair.

Adapted-1 is an equal-size offline control. For each task, it samples
one of the four source-conditioned descriptions and applies it to all
four targets, matching Original in evaluation size while keeping the
description shared across targets. We evaluate 10,000 such assignments.
Generic provides a public-only control: it produces one shared rewrite
per task using the same generator and task-preservation constraints,
but without trajectories, LM--CLI identities, previous outcomes, or
evaluator metadata. Generic is evaluated on all four target pairs,
yielding 64 additional pair--task executions.


\begin{figure}[!t]
  \centering
  \includegraphics[
    width=\linewidth,
    trim=0bp 2bp 0bp 2bp,
    clip
  ]{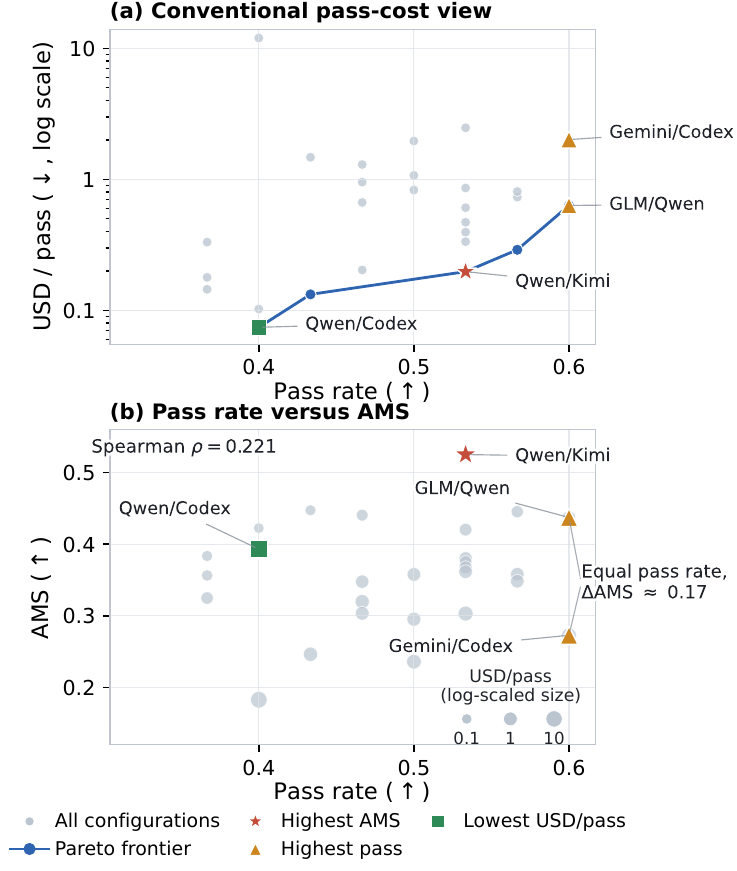}
  \vspace{-2pt}

  \caption{Pass--cost and pass--AMS views of the 28 Core
configurations. Pass--cost leaves multiple competitive choices,
whereas AMS provides a unified ranking. In (b), gray marker size
denotes USD per pass on a log scale, while colored shapes mark
criterion-specific leaders.}
  \label{fig:ams-added-value}
  \vspace{-3pt}
\end{figure}

\begin{table}[!t]
\centering
\footnotesize

\setlength{\tabcolsep}{2.2pt}
\setlength{\aboverulesep}{0.28ex}
\setlength{\belowrulesep}{0.28ex}
\renewcommand{\arraystretch}{1.02}

\begin{tabular*}{\columnwidth}{
  @{\extracolsep{\fill}}
  lccccc
  @{}
}
\toprule
\tablehead{Variant}
& \tablehead{$\rho$}
& \tablehead{$\tau$}
& \tablehead{Top-5}
& \tablehead{Max shift}
& \tablehead{Winner} \\
\midrule
w/o TQ
& 0.895 & 0.751 & 2/5 & 9 & Same \\
w/o AUBQC
& 0.690 & 0.529 & 3/5 & 13 & Changed \\
w/o EFR
& 0.897 & 0.746 & 4/5 & 9 & Same \\
\bottomrule
\end{tabular*}

\caption{AMS component ablation on the Core Subset. All ranking
statistics are measured against full AMS: Top-5 is the overlap between
the two top-five sets, Max shift is the largest absolute rank change,
and Winner indicates whether the leading configuration changes.}
\label{tab:benchmark-diagnostics}
\vspace{-3pt}
\end{table}

\begin{table}[!t]
  \centering
  \bodytablestyle

  \setlength{\tabcolsep}{1.7pt}
  \renewcommand{\arraystretch}{1.05}

  \begin{tabular*}{\columnwidth}{
    @{\extracolsep{\fill}}lrrrr@{}}
    \toprule
    & \multicolumn{3}{c}{AMS$\uparrow$}
    & \\
    \cmidrule(lr){2-4}

    \tablehead{Pair}
    & \tablehead{Orig.}
    & \tablehead{Pooled}
    & \tablehead{Cross-only}
    & \tablehead{A1 Top-1} \\
    \midrule

    Qwen/Kimi
    & 0.353 & 0.470 & 0.457 & 40.2\% \\
    Qwen/Codex
    & 0.250 & 0.457 & 0.411 & 30.3\% \\
    DS/Kimi
    & 0.341 & 0.457 & 0.438 & 19.9\% \\
    GLM/Qwen
    & 0.389 & 0.446 & 0.429 & 9.6\% \\
    \midrule

    \textbf{Macro}
    & \textbf{0.333}
    & \textbf{0.458}
    & \textbf{0.434}
    & -- \\
    \bottomrule
  \end{tabular*}

  \caption{Description controls on \textsc{AgentMeter-Opt}. Pooled
balances description sources, Cross-only excludes the target source,
and Adapted-1 (A1) Top-1 reports the fraction of single-source
assignments in which each pair ranks first.}
  \label{tab:description-dataset-controls}
  \vspace{-3pt}
\end{table}

\subsection{Result Analysis}
\label{sec:result-analysis}

\noindent\textbf{Q1: Does LM--CLI pairing change the ranking?}

Table~\ref{tab:core30-main-results} yields different leaders under
different evaluation objectives. GLM-5.1/Qwen Code and Gemini 3.5
Flash/Codex CLI lead in completion with 18 passes each, whereas
Qwen3.6-Plus/Kimi Code CLI achieves the highest AMS
(\(0.525\)). Qwen3.6-Plus/Codex CLI has the lowest
successful-execution cost, at \(0.48\)M tokens and \(0.074\) USD per
pass. Completion, cost, and quality--efficiency therefore select
different configurations from the same \(7\times4\) grid.

CLI choice also changes the result within a fixed model. For
Qwen3.6-Plus, AMS ranges from \(0.325\) to \(0.525\) across the four
interfaces. Figure~\ref{fig:agentmeter-overview}(b) further shows that
the best CLI varies across language models. Evaluating every model
through one fixed interface would therefore conflate model capability
with LM--CLI compatibility.

\noindent\textbf{Q2: What does AMS add beyond pass and cost?}
\label{sec:benchmark-diagnostics}

As shown in Figure~\ref{fig:ams-added-value}, conventional pass--cost
metrics do not provide a unified judgment, whereas AMS integrates
quality and efficiency into a single ranking. Across the 28 Core
configurations, AMS is only weakly correlated with pass rate
(Spearman \(\rho=0.221\)). For example, GLM-5.1/Qwen Code and
Gemini 3.5 Flash/Codex CLI both pass 18 tasks, but their AMS values
differ by approximately \(0.17\).

The component ablation in Table~\ref{tab:benchmark-diagnostics}
confirms that no single term determines the AMS ranking. Removing
AUBQC changes the winner and reduces rank correlation with full AMS
to \(0.690\), with a maximum shift of 13 positions. Removing TQ or
EFR preserves the winner but still moves some configurations by up to
nine positions. Across the 16 configurations evaluated on both sets,
Core and Full AMS rankings remain strongly correlated
(\(\rho=0.835\)), supporting the Core Subset as a more efficient
setting for evaluating the complete \(7\times4\) configuration grid.

\begin{figure}[!t]
  \centering

  \includegraphics[
    width=\linewidth,
    trim=3bp 96bp 3bp 38bp,
    clip
  ]{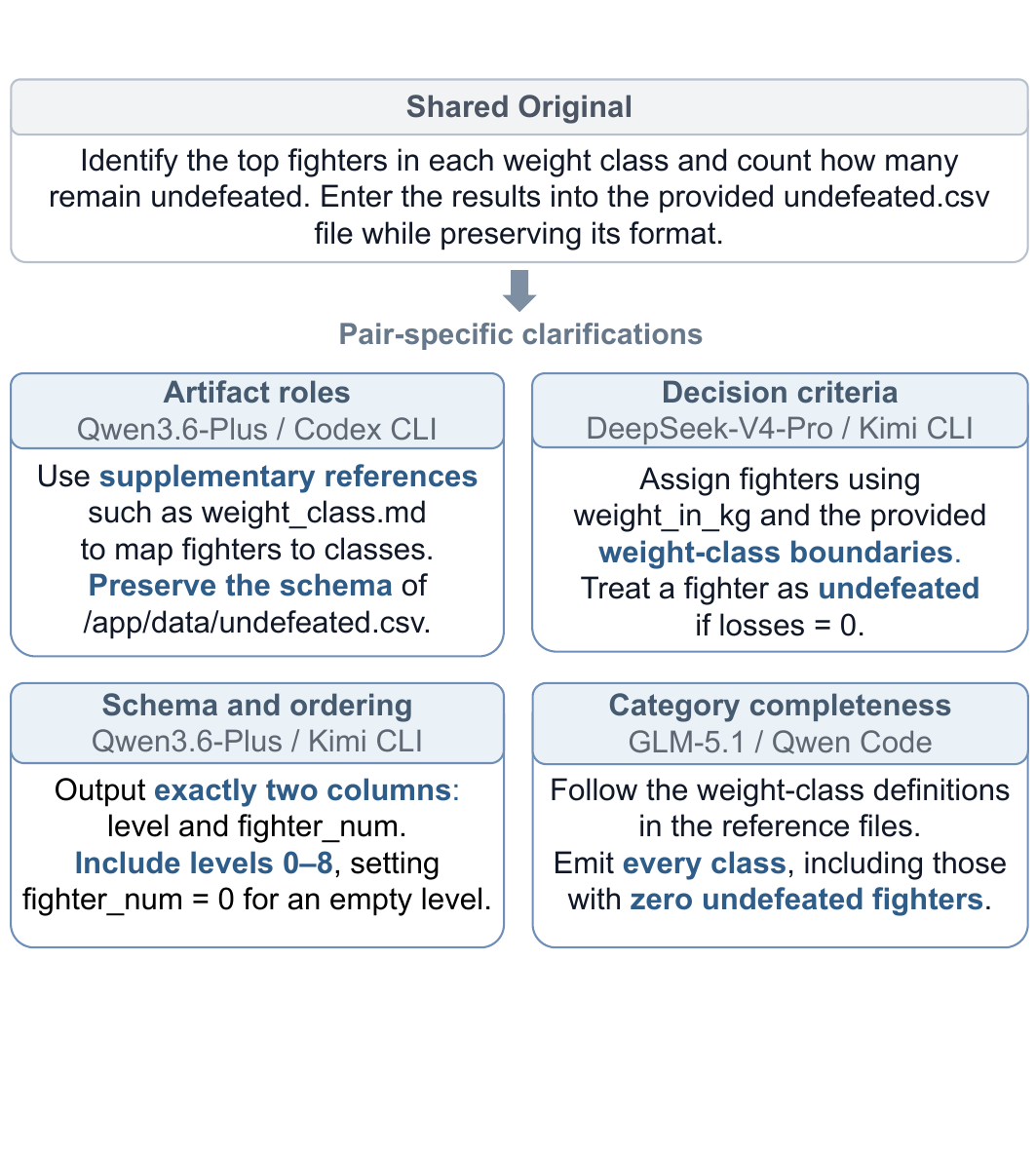}

  \vspace{-3pt}

  \caption{Abridged task-preserving variants for the DA-Code task
\mbox{DM-CSV-001}. Source trajectories expose complementary
ambiguities, with bold text marking the resulting clarifications.}
  \label{fig:qualitative-example}
  \vspace{-3pt}
\end{figure}

\noindent\textbf{Q3: Does pooling reduce description dependence?}
\label{sec:description-sensitivity}

We compare evaluation under the single collected Original description
with single-description and source-balanced alternatives. As a
preservation audit, all 64 retained source-conditioned descriptions
pass the gate, with no recorded task drift, unsupported added
requirements, or fallback to Original. Rejected candidates typically
omit public requirements, alter artifact or path assumptions,
introduce implementation-specific guidance, or reduce task difficulty.

Table~\ref{tab:description-dataset-controls} shows that Pooled raises
macro AMS from \(0.333\) to \(0.458\) and changes the leading
configuration from GLM-5.1/Qwen Code to Qwen3.6-Plus/Kimi Code CLI.
The ranking shift indicates that the Original result depends partly on
the collected formulation. Pooled instead evaluates every target over
the same fixed, source-balanced description distribution for direct
comparison.

Adapted-1 tests whether replacing Original with another single valid
description is sufficient. Across 10,000 equal-size assignments,
Qwen3.6-Plus/Kimi Code CLI ranks first in \(40.2\%\) of cases, whereas
the Original leader, GLM-5.1/Qwen Code, ranks first in only \(9.6\%\).
Moreover, only \(7.2\%\) of the assignments reproduce the exact Pooled
ordering. These results show that evaluation under any single valid
description remains sensitive to which description is selected.
Pooled reduces this source-selection dependence across target
configurations by aggregating multiple source-conditioned descriptions
rather than committing the benchmark to one formulation.

Cross-only removes the target pair's own source description. Its macro
AMS remains \(0.434\), above the Original value of \(0.333\), and all
four pairs retain an improvement over Original. At the condition level,
114 of 192 Cross evaluations either improve reward or preserve reward
at lower cost. The Pooled result therefore does not rely only on
pair-specific Self descriptions. Self is reported as a source-side
diagnostic, whereas Cross-only provides a source-held-out transfer
check.

Figure~\ref{fig:qualitative-example} illustrates the corresponding
task-preserving clarifications for one DA-Code task.

\begin{figure}[!t]
  \centering
  \bodytablestyle

  \setlength{\tabcolsep}{1.6pt}
  \renewcommand{\arraystretch}{1.05}

  \begin{tabular*}{\columnwidth}{
    @{\extracolsep{\fill}}lrrrrrr@{}}
    \toprule
    & \multicolumn{3}{c}{Mean reward$\uparrow$}
    & \multicolumn{3}{c}{AMS$\uparrow$} \\
    \cmidrule(lr){2-4}
    \cmidrule(lr){5-7}

    \tablehead{Pair}
    & \tablehead{Orig.}
    & \tablehead{Generic}
    & \tablehead{Self}
    & \tablehead{Orig.}
    & \tablehead{Generic}
    & \tablehead{Self} \\
    \midrule

    Qwen/Kimi
    & 0.432 & 0.423 & 0.559
    & 0.353 & 0.355 & 0.509 \\

    Qwen/Codex
    & 0.276 & 0.422 & 0.674
    & 0.250 & 0.319 & 0.600 \\

    DS/Kimi
    & 0.552 & 0.521 & 0.708
    & 0.341 & 0.346 & 0.517 \\

    GLM/Qwen
    & 0.521 & 0.583 & 0.646
    & 0.389 & 0.369 & 0.498 \\

    \midrule
    \textbf{Macro}
    & \textbf{0.445}
    & \textbf{0.487}
    & \textbf{0.647}
    & \textbf{0.333}
    & \textbf{0.347}
    & \textbf{0.531} \\
    \bottomrule
  \end{tabular*}

  \vspace{1pt}

  \captionof{table}{Mean reward and AMS under Original, Generic,
  and Self descriptions on \textsc{AgentMeter-Opt}. Generic is
  public-only, whereas Self is trajectory-conditioned.}
  \label{tab:public-only-ablation}
  \vspace{-3pt}
\end{figure}

\noindent\textbf{Q4: Can generic rewriting reproduce the gains?}
\label{sec:generic-rewrite-ablation}

Generic tests whether the large pair-conditioned gains can be
explained by public-only clarification. It uses one shared rewrite per
task and follows the same task-preservation constraints as
\textsc{AgentMeter-Opt}, but has access only to the Original public
instruction.

Table~\ref{tab:public-only-ablation} indicates that the main gains are
not explained by generic public-only rewriting. Generic has
inconsistent effects across pairs and yields only limited aggregate
changes, whereas Self improves both mean reward and AMS over Original
for all four pairs, reaching macro values of \(0.647\) and \(0.531\),
respectively, while reducing aggregate cost by \(37.9\%\). This
contrast suggests that effective description adaptation depends on
diagnosing and clarifying the concrete ambiguities exposed by each
pair's trajectory while preserving the underlying task.


\section{Conclusion}

We introduced \textsc{AgentMeter} to evaluate complete LM--CLI
configurations using AMS, which combines graded task quality, quality
under calibrated cost budgets, and costly zero-reward failures.
Experiments show that pass rate, execution cost, and AMS identify
different leading configurations, and that no CLI is consistently best
across language models. \textsc{AgentMeter-Opt} further shows that
measured performance and ranking can depend on the collected task
formulation. Source-balanced Pooled evaluation reduces dependence on
any formulation, while generic public-only rewriting does not explain
the main gains. This contrast suggests that effective adaptation depends
on diagnosing the concrete ambiguities exposed by each pair's
trajectory. The description analysis focuses on a controlled
source--target setting; broader pair coverage and repeated trials will
help assess the generality of these findings.

\section*{Acknowledgments}
This work was supported by the Strategic Priority Research Program of the
Chinese Academy of Sciences under Grant No. XDA0460205.

\def\agentmeterfullpaper{}
\ifdefined\agentmeterfullpaper
\else
\documentclass[letterpaper]{article}

\usepackage{aaai2026}
\nocopyright
\usepackage{times}
\usepackage{helvet}
\usepackage{courier}
\usepackage[hyphens]{url}
\usepackage{graphicx}
\usepackage{booktabs}
\usepackage{array}
\usepackage{algorithm}
\usepackage{algorithmic}
\usepackage{float}
\usepackage{dblfloatfix}
\usepackage{placeins}
\usepackage{amsmath}
\usepackage{amssymb}

\urlstyle{rm}
\def\UrlFont{\rm}
\frenchspacing
\raggedbottom

\setlength{\textfloatsep}{7pt plus 1pt minus 2pt}
\setlength{\floatsep}{6pt plus 1pt minus 2pt}
\setlength{\intextsep}{6pt plus 1pt minus 2pt}
\setlength{\dbltextfloatsep}{6pt plus 1pt minus 2pt}
\setlength{\dblfloatsep}{6pt plus 1pt minus 2pt}
\setlength{\abovecaptionskip}{5pt}
\setlength{\belowcaptionskip}{3pt}

\setcounter{topnumber}{3}
\setcounter{bottomnumber}{2}
\setcounter{totalnumber}{6}
\setcounter{dbltopnumber}{2}
\renewcommand{\topfraction}{0.98}
\renewcommand{\bottomfraction}{0.90}
\renewcommand{\textfraction}{0.02}
\renewcommand{\floatpagefraction}{0.80}
\renewcommand{\dbltopfraction}{0.98}
\renewcommand{\dblfloatpagefraction}{0.80}

\makeatletter
\setlength{\@fptop}{0pt}
\setlength{\@fpsep}{6pt plus 1pt minus 1pt}
\setlength{\@fpbot}{0pt plus 1fil}
\setlength{\@dblfptop}{0pt}
\setlength{\@dblfpsep}{6pt plus 1pt minus 1pt}
\setlength{\@dblfpbot}{0pt plus 1fil}
\makeatother

\newcommand{\agentmeter}{\textsc{AgentMeter}}
\newcommand{\agentmeteropt}{\textsc{AgentMeter-Opt}}

\title{Supplementary Material for\\
Matching Matters: A Fair Quality--Efficiency Benchmark for Command-Line Agents}

\author{Anonymous Authors}
\affiliations{}

\begin{document}
\maketitle
\fi

\appendix
\setcounter{figure}{0}
\setcounter{table}{0}
\setcounter{equation}{0}
\renewcommand{\thefigure}{A\arabic{figure}}
\renewcommand{\thetable}{A\arabic{table}}
\renewcommand{\theequation}{A\arabic{equation}}
\setcounter{algorithm}{0}
\renewcommand{\thealgorithm}{A\arabic{algorithm}}

\section{Benchmark Construction and Calibration}
\label{app:benchmark-calibration}

\subsection{Observed-Effort and Cost Calibration}

We calibrate the observed-effort index using runs under the Original
descriptions on the Full Benchmark. Equation~(1) provides a compact summary
of the three task-level effort statistics. In the reported implementation,
the failure-related statistic is computed as the non-pass frequency based on
the binary full-pass indicator.

For each effort stratum, the AUBQC budget grid uses the 20th, 40th, 60th,
80th, and 90th percentiles of billable cost among positive-reward reference
executions. The expensive-failure threshold is the 90th percentile of
zero-reward reference cost within the same stratum. The non-pass statistic
is used for task-effort ordering and stratum assignment, whereas EFR is
defined using zero-reward executions. Partial-reward non-pass runs contribute
to TQ and AUBQC but are not counted as expensive zero-reward failures. The
resulting calibration values are reported in Table~\ref{tab:app-budget}.

\begin{table}[!htbp]
\centering
\footnotesize
\setlength{\tabcolsep}{3.8pt}
\renewcommand{\arraystretch}{1.12}
\caption{Tier-specific AUBQC budgets and zero-reward \(P_{90}\)
thresholds in USD.}
\label{tab:app-budget}
\begin{tabular*}{\columnwidth}{@{\extracolsep{\fill}}lrrrrrr@{}}
\toprule
Tier & P20 & P40 & P60 & P80 & P90 & ZR \\
\midrule
Easy   & 0.009 & 0.018 & 0.038 & 0.071 & 0.095 & 0.070 \\
Medium & 0.024 & 0.051 & 0.097 & 0.202 & 0.349 & 0.297 \\
Hard   & 0.112 & 0.280 & 0.458 & 0.822 & 1.455 & 1.057 \\
\bottomrule
\end{tabular*}
\end{table}

\subsection{Evaluation Sets}

We construct three evaluation sets from the same executable-task pool. The
80-task Full Benchmark supplies the calibration reference; the 30-task Core
Subset supports complete evaluation of the \(7\times4\) LM--CLI grid; and the
16-task \agentmeteropt{} subset supports controlled evaluation under
description variation over four target pairs. Table~\ref{tab:app-sets} summarizes their source,
observed-effort, and task-family composition. We organize the tasks into six broad functional
families based on their primary objectives and required operations.

Core is balanced by ordering its selected tasks using the frozen
Full-calibrated effort index and forming \(10/10/10\) Easy/Medium/Hard strata.
The \agentmeteropt{} subset inherits the corresponding Full-Benchmark effort
labels, producing a \(5/7/4\) split. Core and Full contain 840 and 1,280
executed pair--task runs, respectively. Original, Self, and Cross contribute
320 executed conditions on \agentmeteropt{}; Generic adds 64, while Pooled,
Cross-only, and Adapted-1 are computed offline from the executed matrix.

\begin{table}[!htbp]
\centering
\footnotesize
\setlength{\tabcolsep}{4.0pt}
\renewcommand{\arraystretch}{1.12}
\caption{Composition of the Full Benchmark, Core Subset, and
\agentmeteropt{} subset.}
\label{tab:app-sets}
\begin{tabular*}{\columnwidth}{@{\extracolsep{\fill}}llrrr@{}}
\toprule
Group & Item & Full & Core & Opt \\
\midrule
\multicolumn{5}{@{}l}{\textit{Source benchmark}} \\
 & SkillsBench    & 39 & 16 & 7 \\
 & Terminal-Bench & 23 & 9  & 6 \\
 & DA-Code        & 18 & 5  & 3 \\
\addlinespace[2pt]
\multicolumn{5}{@{}l}{\textit{Observed effort}} \\
 & Easy   & 27 & 10 & 5 \\
 & Medium & 26 & 10 & 7 \\
 & Hard   & 27 & 10 & 4 \\
\addlinespace[2pt]
\multicolumn{5}{@{}l}{\textit{Task family}} \\
 & Data \& analytics             & 24 & 6 & 6 \\
 & Software \& systems           & 14 & 1 & 2 \\
 & Docs/media/visualization      & 13 & 4 & 3 \\
 & Planning/control/optimization & 12 & 8 & 2 \\
 & Scientific \& ML compute      & 10 & 7 & 1 \\
 & Security/low-level tooling    & 7  & 4 & 2 \\
\midrule
Total & & 80 & 30 & 16 \\
\bottomrule
\end{tabular*}
\end{table}

\section{Reproducibility Details}
\label{app:implementation-opt}

\subsection{Runtime and Pricing}

We run each CLI with its default inference configuration and do not tune
decoding parameters or inference strength for individual LM--CLI pairs. The
recorded releases are Codex CLI 0.128.0, Qwen Code 0.15.6, Claude Code
2.1.126, and Kimi CLI 1.41.0.

Table~\ref{tab:app-pricing} reports the model prices used for all
cost-sensitive comparisons. Chinese-provider prices are converted using
\(1\,\mathrm{CNY}=0.1471\,\mathrm{USD}\), and cache-read and cache-creation
prices are included when available.

\begin{table}[!htbp]
\centering
\footnotesize
\setlength{\tabcolsep}{3.5pt}
\renewcommand{\arraystretch}{1.12}
\caption{Model pricing used in our experiments, in USD per million
tokens. R/W denote cache read/write; Chinese-provider prices are converted
using \(1\,\mathrm{CNY}=0.1471\,\mathrm{USD}\).}
\label{tab:app-pricing}
\begin{tabular}{@{}lrrrr@{}}
\toprule
Model & In & R & W & Out \\
\midrule
Qwen3.6-Plus      & 0.294 & 0.059 & --    & 1.765 \\
DeepSeek-V4-Pro  & 1.765 & 0.147 & --    & 3.530 \\
GLM-5.1           & 1.177 & 0.235 & --    & 4.119 \\
MiniMax-M2.5      & 0.309 & 0.031 & --    & 1.236 \\
GPT-5.3-Codex     & 1.750 & 0.175 & --    & 14.000 \\
Claude Sonnet 4.6 & 3.000 & 0.300 & 3.750 & 15.000 \\
Gemini 3.5 Flash & 1.500 & 0.150 & --    & 9.000 \\
\bottomrule
\end{tabular}
\end{table}

Agent execution uses a 3,600-second timeout, with setup limited to 900 seconds
when applicable.

Each pair--task condition follows the same execution and aggregation
protocol. Rewards, token usage, and billable cost are recorded at the
execution level before task-level and stratum-level aggregation.

\subsection{\agentmeteropt{} Construction Details}

Algorithm~\ref{alg:app-opt} summarizes the source-side construction and
retention procedure. For each task--source pair, we extract public task
anchors and compress the source trajectory into a small set of evidence
records used to generate candidate descriptions. All candidates are generated
with Qwen3.7-Max under the same generation configuration.

\begin{algorithm}[!t]
\caption{\agentmeteropt{} description construction and source-side retention}
\label{alg:app-opt}
\begin{algorithmic}[1]
\STATE {\bfseries Input:} Original instruction \(I\), source trajectory \(T\),
and source result \(R=(r_0,c_0)\), where \(r_0\) and \(c_0\) denote reward
and billable cost.
\STATE {\bfseries Output:} Frozen retained description \(I^\star\).

\STATE \textbf{Stage 1: Trajectory-Evidence Construction}
\STATE \(A \leftarrow \mathrm{ExtractPublicAnchors}(I)\).
\STATE \(E \leftarrow \mathrm{CompressTrajectory}(T,A)\).
\STATE \(G \leftarrow \mathrm{BuildRewriteGuidance}(E,R)\).

\item[] \hrulefill
\STATE \textbf{Stage 2: Candidate Screening and Source Validation}
\STATE \(C \leftarrow \mathrm{GenerateCandidates}(I,G)\),
\(\mathcal{V}\leftarrow\emptyset\).
\FOR{each candidate \(c\in C\)}
    \STATE \(s \leftarrow \mathrm{SafetyGate}(I,c)\).
    \IF{\(s=\mathrm{pass}\) and \(c\neq I\)}
        \STATE \((r_c,c_c)\leftarrow\mathrm{EvaluateSource}(c)\).
        \STATE Add \((c,r_c,c_c)\) to \(\mathcal{V}\).
    \ELSE
        \STATE Reject \(c\) and record the reason.
    \ENDIF
\ENDFOR

\item[] \hrulefill
\STATE \textbf{Stage 3: Retention and Freezing}
\STATE Keep candidates with \(r_c>r_0\), or \(r_c=r_0\) and \(c_c<c_0\).
\IF{no candidate remains}
    \STATE \(I^\star \leftarrow I\).
\ELSE
    \STATE Select by higher reward, then lower cost.
    \STATE Break remaining ties by the smaller task-preserving edit.
    \STATE \(I^\star \leftarrow\) the selected candidate.
\ENDIF
\STATE Freeze \(I^\star\) before Self, Cross, and Pooled evaluation.
\STATE \textbf{return} \(I^\star\).
\end{algorithmic}
\end{algorithm}

Only task-preserving clarifications are eligible for source-side reruns.
Candidates that alter the task specification or introduce unsupported
information are rejected. Selection uses source-side outcomes only, and the
retained description is frozen before transfer evaluation.

\section{Additional AMS Results}
\label{app:ams-diagnostics}

\subsection{Full-Benchmark Results and Core Validation}

Table~\ref{tab:app-full16} reports the 16 configurations shared by Full
and Core, with both rank columns recomputed within this common set.

\begin{table*}[!t]
\centering
\footnotesize
\setlength{\tabcolsep}{3.8pt}
\renewcommand{\arraystretch}{1.10}
\caption{Full-Benchmark results and common-16 Core comparison. \(r_F\) and
\(r_C\) are ranks within the same 16 configurations. Tok./P and USD/P divide total token usage and total USD cost,
respectively, across all valid executions by the number of full passes.}
\label{tab:app-full16}
\begin{tabular*}{0.98\textwidth}{
@{\extracolsep{\fill}}llrrrrrrrrr@{}}
\toprule
Model & CLI & Pass & Tok./P & USD/P & TQ & AUBQC & EFR & AMS &
\(r_F\) & \(r_C\) \\
\midrule
DeepSeek-V4-Pro   & Claude Code & 40 & 1.31M & 0.946 & 0.535 & 0.098 & 0.150 & 0.314 & 16 & 16 \\
                  & Codex CLI   & 43 & 1.00M & 0.851 & 0.577 & 0.164 & 0.086 & 0.400 & 14 & 14 \\
                  & Kimi CLI    & 50 & 0.86M & 0.677 & 0.655 & 0.234 & 0.099 & 0.454 & 3  & 9  \\
                  & Qwen Code   & 42 & 1.21M & 0.789 & 0.562 & 0.171 & 0.113 & 0.368 & 15 & 15 \\
\midrule
GLM-5.1           & Claude Code & 43 & 1.28M & 0.610 & 0.576 & 0.207 & 0.075 & 0.404 & 13 & 10 \\
                  & Codex CLI   & 45 & 1.01M & 0.476 & 0.576 & 0.270 & 0.050 & 0.428 & 9  & 12 \\
                  & Kimi CLI    & 44 & 0.95M & 0.441 & 0.592 & 0.291 & 0.050 & 0.453 & 4  & 8  \\
                  & Qwen Code   & 47 & 1.26M & 0.532 & 0.613 & 0.246 & 0.063 & 0.444 & 7  & 4  \\
\midrule
MiniMax-M2.5      & Claude Code & 38 & 1.67M & 0.112 & 0.521 & 0.393 & 0.000 & 0.470 & 2  & 2  \\
                  & Codex CLI   & 34 & 0.89M & 0.075 & 0.459 & 0.426 & 0.000 & 0.446 & 6  & 5  \\
                  & Kimi CLI    & 35 & 1.40M & 0.108 & 0.462 & 0.399 & 0.000 & 0.437 & 8  & 7  \\
                  & Qwen Code   & 34 & 1.83M & 0.119 & 0.446 & 0.391 & 0.000 & 0.424 & 10 & 11 \\
\midrule
Qwen3.6-Plus      & Claude Code & 34 & 2.17M & 0.205 & 0.460 & 0.343 & 0.013 & 0.408 & 12 & 13 \\
                  & Codex CLI   & 33 & 0.44M & 0.065 & 0.437 & 0.401 & 0.013 & 0.418 & 11 & 6  \\
                  & Kimi CLI    & 41 & 1.17M & 0.138 & 0.541 & 0.454 & 0.000 & 0.506 & 1  & 1  \\
                  & Qwen Code   & 35 & 2.34M & 0.221 & 0.491 & 0.387 & 0.012 & 0.447 & 5  & 3  \\
\bottomrule
\end{tabular*}
\end{table*}

Core--Full agreement remains strong in the common-16 audit. In the
size-matched subset audit, Core lies in the \(91.4\)--\(95.0\)th percentile
for Spearman agreement, the \(93.8\)--\(96.2\)nd percentile for Kendall
agreement, and the \(90.9\)--\(94.6\)th percentile under maximum rank shift.

\subsection{Expensive Zero-Reward Executions}

Table~\ref{tab:app-expensive-failures} expands the expensive-failure analysis behind
EFR. Although expensive zero-reward executions form a minority of failures,
they account for a disproportionate share of zero-reward cost. The pair-level
rows identify the configurations most affected by removing EFR.

\begin{table}[!t]
\centering
\footnotesize
\setlength{\tabcolsep}{3.5pt}
\renewcommand{\arraystretch}{1.12}
\caption{Expensive zero-reward executions and EFR effects. Full uses
the 16 configurations with complete coverage; Core and pair-level statistics
use all 28 Core configurations. Exp.\ ZR denotes zero-reward executions whose
cost exceeds the tier-specific zero-reward \(P_{90}\) threshold. Cost share is
the fraction of total zero-reward cost attributable to these executions.
\(\Delta\mathrm{AMS}=\mathrm{AMS}_{\mathrm{w/o\ EFR}}
-\mathrm{AMS}_{\mathrm{default}}\).}
\label{tab:app-expensive-failures}
\begin{tabular*}{\columnwidth}{@{\extracolsep{\fill}}lrrrr@{}}
\toprule
\multicolumn{5}{@{}l}{\textit{(a) Dataset-level concentration}} \\
Dataset & ZR & Exp. ZR & Exp. rate & Cost share \\
\midrule
Full & 563 & 58 & 10.3\% & 48.6\% \\
Core & 373 & 96 & 25.7\% & 79.3\% \\
\addlinespace[3pt]
\multicolumn{5}{@{}l}{\textit{(b) Largest pair-level EFR effects on Core}} \\
Pair & ZR & Exp. ZR & Cost share & \(\Delta\)AMS \\
\midrule
Gemini/Codex & 10 & 9 & 95.0\% & +0.110 \\
Claude/Qwen  & 10 & 8 & 94.7\% & +0.098 \\
Gemini/Qwen  & 13 & 8 & 72.2\% & +0.088 \\
Claude/Codex & 11 & 5 & 82.1\% & +0.069 \\
\bottomrule
\end{tabular*}
\end{table}

This concentration motivates treating EFR separately from budgeted quality:
otherwise, costly zero-reward executions can be obscured by aggregate
budgeted-quality behavior.

\subsection{Parameter Sensitivity}

In all reported experiments, the three task-level statistics summarized in
Equation~(1) of the main paper are aggregated using fixed weights
\((0.50,0.25,0.25)\) for token usage, execution steps, and non-pass
frequency, respectively. The three alternative settings
\((1/3,1/3,1/3)\), \((0.40,0.30,0.30)\), and
\((0.25,0.25,0.50)\) move 4, 2, and 10 tasks across effort strata,
respectively, but retain a \(5/5\) top-five overlap and Spearman correlation
of at least \(0.980\). Varying \(\alpha\) over
\(\{0.4,0.5,0.7,0.8\}\) changes 12--20 ranks, but Qwen3.6-Plus/Kimi remains
the leading configuration; \(\alpha\in\{0.5,0.7\}\) keeps Spearman
correlation above \(0.96\).

\section{Additional Description Diagnostics}
\label{app:description-results}

This section reports task-level and source--target diagnostics omitted from
the main paper.

\subsection{Task-Level and Adapted-1 Controls}

Table~\ref{tab:app-description-controls} reports task-level reward changes
and assignment-level variation for the same-size Adapted-1 control.

\begin{table}[!t]
\centering
\footnotesize
\caption{Task-level reward changes and Adapted-1 assignment variation. Panel
(a) reports task-level paired-bootstrap 95\% intervals; panel (b) reports the
central 95\% range across 10,000 equal-size assignments.}
\label{tab:app-description-controls}

\textit{(a) Pair--task reward changes}\\[3pt]
\setlength{\tabcolsep}{2.5pt}
\renewcommand{\arraystretch}{1.12}
\begin{tabular*}{\columnwidth}{@{\extracolsep{\fill}}lrrrrl@{}}
\toprule
Condition & Better & Same & Worse & Mean \(\Delta R\) & 95\% CI \\
\midrule
Self   & 16 & 48 & 0 & \(+0.2015\) & [0.0938, 0.3196] \\
Pooled & 22 & 35 & 7 & \(+0.1379\) & [0.0690, 0.2109] \\
\bottomrule
\end{tabular*}

\vspace{5pt}
\textit{(b) Adapted-1 assignment variation}\\[3pt]
\setlength{\tabcolsep}{3.0pt}
\begin{tabular*}{\columnwidth}{@{\extracolsep{\fill}}lrrr@{}}
\toprule
Target & Mean AMS & 95\% range & Top-1 prob. \\
\midrule
Qwen/Kimi  & 0.471 & [0.360, 0.598] & 40.2\% \\
Qwen/Codex & 0.457 & [0.326, 0.577] & 30.3\% \\
DS/Kimi    & 0.458 & [0.346, 0.570] & 19.9\% \\
GLM/Qwen   & 0.446 & [0.345, 0.539] & 9.6\% \\
\bottomrule
\end{tabular*}
\end{table}

Self has no task-level reward decreases by construction: a candidate is
retained only if it improves source reward or preserves reward at lower cost,
and the Original description is used otherwise. In contrast, Adapted-1 exhibits broad
assignment-level variation, indicating that a single sampled description does
not reliably reproduce the source-balanced result.

\subsection{Cross-Transfer Directionality}

Of the 192 off-diagonal Cross evaluations, 114 (\(59.4\%\)) improve reward or
preserve reward at lower cost. Table~\ref{tab:app-cross-direction} reports the
effectiveness rate for each source--target direction.

\begin{table}[!t]
\centering
\footnotesize
\setlength{\tabcolsep}{4.0pt}
\renewcommand{\arraystretch}{1.10}
\caption{Cross-transfer effectiveness rate (\%) by description source
(rows) and evaluation target (columns). Q/K, Q/C, DS/K, and GLM/Q denote
Qwen/Kimi, Qwen/Codex, DeepSeek/Kimi, and GLM/Qwen, respectively. Diagonal
Self conditions are omitted.}
\label{tab:app-cross-direction}
\begin{tabular*}{\columnwidth}{@{\extracolsep{\fill}}lrrrr@{}}
\toprule
Source \(\backslash\) Target & Q/K & Q/C & DS/K & GLM/Q \\
\midrule
Qwen/Kimi  & --   & 43.8 & 68.8 & 62.5 \\
Qwen/Codex & 62.5 & --   & 75.0 & 62.5 \\
DS/Kimi    & 62.5 & 37.5 & --   & 68.8 \\
GLM/Qwen   & 68.8 & 43.8 & 56.2 & --   \\
\bottomrule
\end{tabular*}
\end{table}

Cross effectiveness ranges from \(37.5\%\) to \(75.0\%\), showing that
transfer is direction-dependent. Transfers between pairs sharing Kimi CLI are
effective in \(65.6\%\) of cases, compared with \(53.1\%\) for the same model
across Kimi and Codex.

\begin{table*}[!t]
\centering
\small
\setlength{\tabcolsep}{3.8pt}
\renewcommand{\arraystretch}{1.08}
\caption{Instruction-level examples of \agentmeteropt{} construction,
transfer, and gate rejection.}
\label{tab:app-qualitative}
\begin{tabular}{@{}
>{\raggedright\arraybackslash}p{0.15\textwidth}
>{\raggedright\arraybackslash}p{0.24\textwidth}
>{\raggedright\arraybackslash}p{0.35\textwidth}
>{\raggedright\arraybackslash}p{0.20\textwidth}
@{}}
\toprule
Case & Original excerpt & Retained clarification or rejected edit & Decision and outcome \\
\midrule
\textbf{Effective Cross}\par
\textit{lru-cache-with-ttl}\par
GLM/Q \(\rightarrow\) Q/K
&
TTL begins on insertion/update; \texttt{get} updates LRU order; eviction
removes the least-recently-used non-expired entry.
&
Specifies TTL reset on update, missing/expired-key LRU behavior, deletion
semantics, and the new-key eviction boundary.
&
\textbf{Accepted.}\par
Source \(R:0.0\rightarrow0.0\), cost \(-37.7\%\).\par
Target \(R:0.0\rightarrow1.0\), cost \(-64.5\%\).
\\
\addlinespace[3pt]
\textbf{Negative Cross}\par
\textit{sales-pivot-analysis}\par
GLM/Q \(\rightarrow\) DS/K
&
Read the public PDF/XLSX inputs and produce the required five-sheet workbook.
&
Specifies input roles, repeated PDF headers, the public merge key, workbook
structure, quartile labels, and income calculation.
&
\textbf{Accepted.}\par
Source \(R:0.0\rightarrow1.0\), cost \(-2.7\%\).\par
Target \(R:1.0\rightarrow0.0\), cost \(-49.0\%\).
\\
\addlinespace[3pt]
\textbf{Rejected by gate}\par
\textit{regex-log}\par
Q/C
&
Save the regex to \texttt{/app/regex.txt}; the public example reads this file.
&
The candidate adds completion text but removes the public file-reading example.
&
\textbf{Rejected before rerun.}\par
Dropped public requirement.
\\
\bottomrule
\end{tabular}
\end{table*}

\subsection{Qualitative Audit}

Table~\ref{tab:app-qualitative} gives compact examples of accepted transfer,
negative transfer, and gate rejection. Negative transfer passes the
task-preservation gate but harms a target, whereas a rejected edit is
discarded before rerun.

\section{Further Discussion and Limitations}
\label{app:discussion}

\noindent\textbf{Evaluation unit and interface attribution.}
The observed variation across CLIs should not be interpreted as a causal
estimate of any single interface mechanism or as evidence that one CLI is
universally superior. A CLI bundles its prompting protocol, tool schema,
observation formatting, context management, and failure-recovery behavior,
which jointly shape the execution trajectory. We therefore treat the complete
LM--CLI configuration as the evaluation unit while holding tasks, execution
environments, and evaluators fixed. Our results show that conclusions obtained
under one interface need not transfer to another interface for the same
language model. Isolating the contribution of individual CLI components would
require controlled reimplementations that vary one mechanism at a time. Such
an analysis would complement this benchmark, but addresses a different
causal question from comparing deployable LM--CLI configurations.

\noindent\textbf{Interpretation and scope of AMS.}
AMS is intended as a deployment-oriented comparison criterion rather than a
universal utility function. Its components answer distinct questions: TQ
measures the absolute task quality obtained, AUBQC measures how much of that
quality remains available under calibrated cost budgets, and EFR distinguishes
early failures from unusually costly zero-reward executions. Reporting only
TQ would treat inexpensive and expensive executions equally, whereas reporting
only budget-conditioned quality could favor inexpensive but low-quality
partial progress. We therefore report pass and cost statistics alongside AMS
rather than claiming that AMS replaces every application-specific objective.
The component ablation shows that the ranking is not determined by one term,
while the parameter-sensitivity analysis preserves the leading configuration
under the tested settings. Nevertheless, deployments with substantially
different quality--cost preferences may reasonably select a different
configuration.

\noindent\textbf{Source adaptation and the fairness claim.}
Because a Self description is selected using outcomes from its source pair,
improvement under Self is expected and should not by itself be interpreted as
evidence of a fairer ranking. We use Self as a source-side diagnostic of
whether the observed trajectory exposes actionable description friction. The fairness analysis instead relies on evaluation conditions that are fixed
before target evaluation and independent of target-side outcomes.
Each retained description is frozen before transfer, no target-side
trajectory, reward, or cost is used to select or revise it, and every target is
evaluated on the same tasks and the same four source-conditioned descriptions
with equal source weights. Cross-only further excludes the description derived
from the target pair itself. Thus, Pooled reduces dependence on the arbitrary
choice of one collected formulation rather than rewarding each target using a
description optimized from its own outcomes.

This construction does not eliminate every possible source of benchmark bias.
In particular, Pooled represents a controlled, source-balanced description
distribution rather than the natural distribution of instructions written by
real users. Estimating that distribution would require a separate collection
study involving multiple independent human formulations. Accordingly,
\agentmeteropt{} should be interpreted as a controlled description-sensitivity
analysis and source-balanced evaluation protocol, not as a complete model of
real-world instruction variation.

\noindent\textbf{Relation to prompt optimization.}
\agentmeteropt{} shares candidate generation and execution-based selection
mechanisms with automatic prompt optimization, but uses them for a different
purpose. Conventional prompt optimization typically adapts a prompt for the
same target system on which the optimized prompt will be deployed. Here, each
description is selected using source-side evidence, frozen, and subsequently
evaluated on both its source and other targets. Cross, Cross-only, Adapted-1,
and Pooled therefore examine transfer and dependence on description selection,
rather than providing additional target-side tuning opportunities.

\noindent\textbf{Dependence on the rewrite generator.}
Using one rewrite generator and one generation configuration holds the
construction mechanism fixed across all source pairs. The Generic control
further separates the effect of generic public-only clarification from that
of trajectory-conditioned adaptation: applying the same generator without
trajectory evidence produces smaller and less consistent changes. This
comparison suggests that the reported effects are not explained solely by the
generator rewriting the public instructions. Nevertheless, another generator,
a different candidate-search procedure, or a collection of human-written
reformulations could produce a different description set. Our results therefore
characterize the controlled construction studied here rather than every
possible task-preserving rewrite procedure.
\subsection{Limitations}

The reported rankings are conditional on the selected tasks, calibration
reference, pricing configuration, and execution harness. The Core--Full audit supports Core as an efficient approximation within
the studied benchmark pool, while the 16-task, four-pair
\agentmeteropt{} experiment is a controlled diagnostic rather than an
exhaustive estimate over all models, interfaces, and domains. The tested
sensitivity settings preserve the leading configuration but do not cover
every possible quality--cost objective. The preservation gate rejects explicit
changes to public task semantics but cannot formally prove natural-language
equivalence. Task-level bootstrap intervals quantify variation over tasks,
and reported USD values exclude local
infrastructure costs.

\ifdefined\agentmeterfullpaper
\else
\end{document}
\fi

\bibliography{references}

\end{document}